\documentclass[preprint]{imsart}

\def\ps@imsheadings{%
	\def\@oddfoot{\hfill\info@line}%
	\let\@evenfoot\@oddfoot%
	\def\@evenhead{\runninghead@size\hfill\leftmark/\rightmark\hfill\llap{\pagenumber@size\thepage}}%
	\def\@oddhead{\runninghead@size\hfill\leftmark/\rightmark\hfill\llap{\pagenumber@size\thepage}}}

\RequirePackage{amsthm,amsmath,amsfonts,amssymb}
\RequirePackage[authoryear]{natbib}
\RequirePackage{graphicx}

\startlocaldefs

\theoremstyle{remark}

\newcommand{\NN}{\mathbb{N}}

\newcommand{\RR}{\mathbb{R}}

\DeclareMathOperator{\EE}{\mathbb{E}}
\DeclareMathOperator{\PP}{\mathbb{P}}
\DeclareMathOperator{\Var}{Var}

\endlocaldefs

\begin{document}

\begin{frontmatter}
\title{Spatial Modeling of Heavy Precipitation by Coupling Weather Station Recordings and Ensemble Forecasts with Max-Stable Processes}

\runtitle{}

\begin{aug}
\author[A]{\fnms{Marco} \snm{Oesting}\ead[label=e1]{oesting@mathematik.uni-siegen.de}},
\and\author[B]{\fnms{Philippe} \snm{Naveau}\ead[label=e2]{philippe.naveau@lsce.ipsl.fr}}

\address[A]{
Department Mathematik, University of Siegen,
\printead{e1}}

\address[B]{Laboratoire de Sciences du Climat et de l'Environnement, IPSL-CNRS,
\printead{e2}}
\end{aug}

\begin{abstract}
Due to complex physical  phenomena, the  distribution of heavy rainfall events is difficult to model spatially. Physically based numerical models can often provide physically coherent spatial patterns, but may miss some important precipitation  features like heavy rainfall intensities. Measurements at ground-based weather stations, however, supply adequate rainfall intensities, but most national weather recording networks are often spatially too sparse to capture rainfall patterns adequately. To bring the best out of these two sources of information, climatologists and hydrologists have been seeking  models that can efficiently  merge different types of rainfall data. One  inherent difficulty is to capture the appropriate  multivariate dependence structure among rainfall maxima. 
For this purpose, multivariate extreme value theory suggests the use of a max-stable process. Such a process can be represented by a max-linear combination of independent copies of a hidden stochastic process weighted by a Poisson point process. In practice, the choice of this hidden process is non-trivial, especially if anisotropy, non-stationarity and nugget effects are present in the spatial data at hand.  
By coupling forecast ensemble data from the French national weather service (M\'et\'eo-France) with local observations, we construct and compare  different types of data driven max-stable processes that are parsimonious in  parameters, easy to simulate and capable of reproducing nugget effects and spatial non-stationarities. We also compare our new method with classical approaches from spatial extreme value theory such as Brown--Resnick processes.
\end{abstract}


\end{frontmatter}

%%%%%%%%%%%%%%%%%%%%%%%%%%%%%%%%%%%%
\section{Introduction}\label{sec:intro} 
%%%%%%%%%%%%%%%%%%%%%%%%%%%%%%%%%%%%
A recent review report on weather and climate extremes \cite{chen-etal-18} 
has emphasized the need of ``theoretical frameworks and statistical methods" for modeling complex extreme events. This brings  the particular question of how to statistically model the dependence among extremes. The field of multivariate extreme value theory (EVT), see the books by \cite{resnick-08, EKM-97, DHF-06}, for instance, offers a particular mathematical framework to address such a question. 

In this work, we deal with modelling extreme precipitation.  
There exists a series of articles that have developed parametric EVT models to capture the multivariate extremal dependence structure of daily or subdaily precipitation \cite[see, e.g.][]{CNN-07, keef-etal-09, bechler-etal-15, buhl-klueppelberg-16, SSTK-17, defondeville-davison-18}.
Keeping in mind that heavy rainfall events are difficult to model due to their large spatial and temporal variability  \cite[see, e.g.][]{sillmann-etal-17}, 
the choice of a parametric model that adequately characterizes rainfall heterogeneity is a non-trivial problem. For example, heavy rainfall events in the south of France are modulated by various effects due to the French orography, see the bottom panel in Figure \ref{fig:grid}, and different types of atmospheric patterns \cite[see][for instance]{CNAC-12, CNN-17, BMT-18}. 
These effects raise the question if relevant spatial information about rainfall patterns can be found to complement observations measured by weather stations. 
This inquiry  moves the focus from  finding and fitting complex parametric multivariate EVT structures   to the problem of coupling local precipitation records with data at different spatial scales.   
  
Nowadays, numerical weather models based on assimilation schemes are able to
provide accurate ensemble forecasts of various atmospheric variables, in
particular temperature fields \cite[see, e.g.][]{TMZN-16}. Due to their large
variability, heavy rainfall intensities remain, even in terms of their 
probability distributions, a challenge for weather forecasts. In particular,
forecasted rainfall extremes may strongly differ from precipitation records
measured at weather stations. The later provide high quality data at the local 
spatial scale (the weather station), but high quality and well maintained 
observational networks have a spatial resolution which is much worse than
current ensemble forecasts. To illustrate this discrepancy of spatial scales
between station networks and weather numerical models, the top panel of Figure
\ref{fig:grid} superimposes the well monitored 110 M\'et\'eo France weather 
station locations on top of the PEARP grid \cite[see][for instance]{TMZN-16} 
which is used for the numerical weather prediction system operated by M\'et\'eo 
France.

Despite the challenges in modeling heavy precipitation intensities accurately, ensemble forecasts of rainfall data still provide relevant information in terms of spatial rainfall patterns, \cite[see][for instance]{Taillardat2019}. Compared to climate models, 
numerical weather models contain fine scale features and have complex 
parametrizations, and throughout their assimilation schemes, they spatially 
track  storm patterns. This makes them good candidates as proxies of 
realistic rainfall patterns although their intensities can be misrepresented.  

In this context, our main question is how to couple ensembles of rainfall forecasts within the construction of models suggested by multivariate EVT in order to simulate coherent spatial fields of extreme precipitation, while preserving the spatial structure observed in weather stations. As a direct by-product, any solution of our research problem is supposed to provide extreme precipitation fields at a very fine scale, the one of the ensemble forecast, see Figure \ref{fig:grid} (top).
To reach these objectives, Section \ref{sec:intro-max}  provides some theoretical background on the underlying statistical models used in our further analysis, i.e.\ max-stable models from EVT. 
Our two available precipitation datasets, observational records at weather stations in the south of France and forecast ensembles on a grid, are presented in Section \ref{sec:data}. 
In Section \ref{sec:models}, we present different ways to couple both types of data within a max-stable framework. 
More precisely,  four data driven models are introduced, fitted to the region of interest, see Figure \ref{fig:grid},   and also compared to  a classical stationary spatial max-stable model, the Brown--Resnick process.
The paper closes with a discussion in Section \ref{sec:conclusion}.

%%%%%%%%%%%%%%%%%%%%%%%%%%%%%%%%%%%%
\section{Max-stable models}\label{sec:intro-max} 
%%%%%%%%%%%%%%%%%%%%%%%%%%%%%%%%%%%%

In this section, we will provide the theoretical background on the models
we will use to model heavy precipitation events.

We start with one of the main results from univariate extreme value theory,
the Fisher--Tippett Theorem \citep{fisher-tippett-28}. Assume that we have
independently and identically distributed random variables $X_1, X_2, \ldots$,
e.g.\ different precipitation measurements at the same station or different forecasts 
for the same grid cell. If there exist normalizing sequences $a_n > 0$ and 
$b_n \in \RR$, $n \in \NN$, such that the normalized maximum
$ M_n = a_n^{-1} (\max_{i=1,\ldots,n} X_i - b_n) $
converges in distribution, i.e.\
$$ \PP\left( a_n^{-1} \left( \max_{i=1,\ldots,n} X_i - b_n \right) \leq x\right)
\stackrel{n \to \infty}{\longrightarrow} G(x), \qquad x \in \RR, $$
for some non-degenerate limit distribution $G$, then $G$ is necessarily a
Generalized Extreme Value (GEV) distribution
$$ 
G(x) = G_{\mu, \sigma, \xi}(x) = 
\exp\bigg( - \left\{ 1 + \xi \frac{x-\mu}{\sigma}\right\}^{-1/\xi}_+\bigg), \qquad x \in \RR, 
$$
for a location $\mu \in \RR$, a scale $\sigma>0$ and a shape parameter 
$\xi \in \RR$. Therefore, the distribution of maxima over certain time periods 
(so-called blocks) at a single station or grid cell are typically modeled by a 
GEV distribution. Note that the result above still holds true if the observations
$X_1, X_2, \ldots$ are not independent, but satisfy certain mixing conditions \cite[see][]{LLR-83}.

For modeling extreme events in space, we need an extension of the above limit result to stochastic processes: Let $X_1(\cdot), X_2(\cdot), \ldots$ be independent copies of a stochastic process $\{X(s): \ s \in S\}$ on some countable index set $S \subset \RR^d$, e.g.\ a dense grid. Then, provided that there exist sequences of normalizing functions $a_n(s) > 0$ and $b_n(s) \in \RR$, 
$s \in S$, $n \in \NN$, such that the process of normalized pointwise maxima
$$M_n(s) = a_n(s)^{-1} \left(\max_{i=1,\ldots,n} X_i(s) - b_n(s)\right)$$ converges in distribution to a stochastic process 
$\{Y(s): \ s \in S\}$ with non-degenerate marginal distributions, this process
is necessarily max-stable.
From the univariate result, we immediately obtain that $Y(s)$ follows a GEV
distribution $Y(s) \sim G_{\mu(s),\sigma(s),\xi(s)}$, $s \in S$. 

In this paper, we focus on the extremal spatial dependence structure. Therefore, by appropriate marginal transformations, we assume that $Y$ possesses unit Fr\'echet margins, i.e.\ $Y(s) \sim G_{1,1,1}$ for all $s \in S$. Such a process is called simple max-stable. By \cite{dehaan-84}, every simple max-stable process can be represented as
\begin{equation} \label{eq:spec-repr}
 Y(s) = \max_{i \in \NN} A_i Z_i(s), \qquad s \in S,
\end{equation}  
where $\{A_i\}_{i \in \NN}$ are the points of a Poisson point process on $(0,\infty)$ with intensity $a^{-2} \mathrm{d}a$ and, independently from the Poisson points, $Z_i$, $i \in \NN$, are independent copies of a stochastic process $\{Z(s), \ s \in S\}$ such that $\EE Z(s) = 1$
for all $s \in S$.

As the intensity of the Poisson point process is fixed for given marginal
distributions, the spatial dependence structure is fully determined by the 
multivariate distributions of the so-called spectral process $Z$. Many classes of max-stable models 
are given by specific choices of $Z$. For instance, a process of the form
$$ Z(s) = \exp\left(W(s) - \frac 1 2 \Var(W(s))\right), \quad s\in S,$$
for some centered Gaussian process $W$, leads to a so-called Brown--Resnick process -- one of the most popular max-stable models. If $S$ is a grid and $W$ has stationary increments, the corresponding Brown--Resnick process $Y$ is stationary and its law depends on the variogram
\begin{equation} \label{eq:vario}
\gamma_W(h) := \frac 1 2 \EE \left[ \left(W(s+h) - W(s)\right)^2 \right], \qquad h \in S, 
\end{equation}
only. Therefore, $Y$ is also called Brown--Resnick process associated to variogram $\gamma_W$ \citep{brown-resnick-77, KSH-09}.

If the spectral process follows a discrete uniform distribution on some finite
set of nonnegative functions $\{z_1,\ldots,z_N\}$ on $S$ with
$\frac 1 N \sum_{i=1}^N z_i(s) = 1$ for all $s \in S$, representation \eqref{eq:spec-repr} simplifies to the max-linear model \citep{wang-stoev-2011}
\begin{equation} \label{eq:maxlin}
 Y(s) = \frac 1 N \max_{i=1}^N A_i^* z_i(s), \qquad s \in S,
\end{equation}
where $A_1^*,\ldots,A_N^*$ are independent unit Fr\'echet random variables.

As an alternative to the max-linear model \eqref{eq:maxlin} which is given by the maximum over a finite number of basis functions $z_1,\ldots,z_N$, \cite{reich-shaby-12} developed a max-stable model that can be written as a sum of
$z_1,\ldots,z_N$, see also \cite{oesting-18} for a further generalization:
\begin{equation} \label{eq:reich-shaby}
 Y(s) = \frac 1 N U(s) \cdot \left( \sum\nolimits_{i=1}^N B_i z_i^{1/\alpha}(s) \right)^\alpha, \qquad s \in S, 
\end{equation}
where $\{U(s), \ s \in S\}$ is a noise process with $1/\alpha$-Fr\'echet
marginal distributions and $B_1,\ldots,B_N$ are i.i.d.\ $\alpha$-stable
random variables whose distribution is given by the Laplace transform
\begin{equation*}
 \EE\left[ \exp(-t B_i) \right] = \exp(-t^\alpha),\qquad t > 0, \ i=1,\ldots,N,
\end{equation*}
for some $\alpha \in (0,1)$. Compared to the max-linear model \eqref{eq:maxlin}, the basis functions in the \cite{reich-shaby-12} allow a multiplicative random nugget effect in model \eqref{eq:reich-shaby}.
This nugget effect and the simple additive form  based on a finite sum make this model attractive as a spatial model in environmental applications.

A popular dependence measure for a max-stable process $Y$ with marginal distribution $G$ is the
extremal coefficient $\theta(s_1,s_2) \in [1,2] $ defined via
\begin{equation} \label{eq:def-ec}
\PP( G(Y(s_1)) \leq u, G(Y(s_2)) \leq u) = \PP( G(Y(s_1)) \leq u)^{\theta(s_1,s_2)}, \qquad u \in [0,1].
\end{equation}
Note that, here, $\theta(s_1,s_2)$ does not depend on $u \in [0,1]$ 
and, if $Y(s_1) = Y(s_2)$ a.s.,  then $\theta(s_1,s_2) = 1$, while $\theta(s_1,s_2) = 2$ corresponds to the independence case.
 For a simple max-stable process with general representation \eqref{eq:spec-repr}, the extremal coefficient can also be expressed as 
$$ \theta(s_1, s_2) = \EE \left[ \max\{Z(s_1), Z(s_2)\} \right]. $$ 
This allows us to make the link with  max-stable models studied in this work,
$$
 \theta(s_1,s_2) = \left\{ 
	\begin{array}{ll}
		\frac 1 N \sum\nolimits_{i=1}^N \max\{z_i(s_1), z_i(s_2)\} & \mbox{, for the max-linear model \eqref{eq:maxlin},} \\
		\frac 1 N \sum\nolimits_{i=1}^N (z_i(s_1)^{1/\alpha} + z_i(s_2)^{1/\alpha})^{\alpha}  & \mbox{, for  model \eqref{eq:reich-shaby},}\\
		2 \Phi\left(\sqrt{\frac{\gamma_W(s_1-s_2)}{2}}\right)  & \mbox{, for a Brown--Resnick process,}
	\end{array}
	\right.
$$
where $\Phi$ denotes the standard normal distribution function. 

In classical geostatistics, spatial dependence is often summarized via variograms which correspond to $L_2$-distances, see Equation \eqref{eq:vario}. 
As both variance and expectation are non-finite in case of unit Fr\'echet margins, other distances have to be used in such cases. For example, \cite{CNP-06} studied a marginal free $L_1$ distance, called a $F$-madogram, 
$$ \frac{1}{2} \mathbb{E} \left| G(Y(s_1))-G(Y(s_2)) \right|, \qquad s_1, s_2 \in S.$$
To understand strong local dependences from a geostatistician perspective, one can notice  that 
the extremal coefficient and the $F$-madogram of a max-stable process are  related 
via
 \begin{equation} \label{eq:extr-coeff}
 \theta(s_1,s_2)  = \frac{1+\mathbb{E} \left| G(Y(s_1))-G(Y(s_2)) \right|}{1-  \mathbb{E} \left| G(Y(s_1))-G(Y(s_2)) \right|}.
 \end{equation} 
%In addition,  relation \eqref{eq:extr-coeff} 
This implies that the spatial structure in any max-stable process $Y$
 can be related to the spatial structure of the input $Z$. More precisely, 
  the madogram of the  spectral process in \eqref{eq:spec-repr} can be linked to the F-madogram by
  \begin{equation} \label{eq:mado}
 \frac{1}{2} \mathbb{E} | Z(s_1) -  Z(s_2)| =  \frac{2 \;  \mathbb{E} | G(Y(s_1)) - G(Y(s_2))| }{1- \mathbb{E} | G(Y(s_1)) - G(Y(s_2))|}.
 \end{equation}
This one-to-one link between the generative input $Z$ and the output $Y$ leads, in the absence of any nugget effect,  to 
$$
\lim_{\|s_1 - s_2\| \to 0} \frac{\mathbb{E}|Z(s_1) - Z(s_2)|}{2 \mathbb{E} \left| G(Y(s_1))-G(Y(s_2)) \right|} =2. 
$$
Keeping in mind that $\EE Z(s) = 2 \EE G(Y(s_1)) =1 $, this limiting result implies that 
distance between the distributions at two nearby locations in the input process $Z(s)$  becomes twice smaller in the output process $2G(Y(s))$.  
So, creating strong extremal dependences implies the need of strong dependence in the generating process. On the contrary, a nugget effect, i.e.\ imperfect dependence even at infinitesimal distances, may appear in the output process only if it is present in the input process.

Still, Equality (\ref{eq:extr-coeff}) tells us that the extremal coefficient $\theta(s_1,s_2)$  will be ideally modeled if and only if  $\mathbb{E} \left| G(Y(s_1))-G(Y(s_2)) \right|$ is well captured. The later condition can only be satisfied if a very similar dependence in $Z(s)$ is built in, see relation (\ref{eq:mado}).  
This reasoning leads to our main modeling idea. Instead of building complex parametric models for $Z(s)$ with inference schemes that typically result in high-dimensional optimization problems, see, for instance, \citet{PRS10, DEO17, HDRG19}, for likelihood-based inference methods or \citet{OSF-17,EKS-18} for (weighted) least square fits of certain summary statistics, we can ``just" plug forecast ensemble members as  max-stable constructions like the ones defined by (\ref{eq:spec-repr}), (\ref{eq:maxlin}) or (\ref{eq:reich-shaby}).   
Equalities (\ref{eq:extr-coeff}) and (\ref{eq:mado}) indicate that, if a subset of  ensemble forecast members is well-chosen, then extremal coefficients measured from the weather stations should be well reproduced. By construction, such a model is based on a very small number of parameters only, and, thus, is easy to fit and simulate.

Before closing this section, we can note that extremal coefficients and madograms only provide specific information about pairwise dependences and do not capture 
multivariate features. Still, the same ideas could be extended to multivariate versions  and complete dependence \cite[see][]{marcon-etal-17,NGCD-09}.  
%structure as spectral process for the construction of a max-stable process.

%%%%%%%%%%%%%%%%%%%%%%%%%%%%%%%%%%%%
\section{Description of Rainfall Data}\label{sec:data} 
%%%%%%%%%%%%%%%%%%%%%%%%%%%%%%%%%%%%
  
\begin{figure}[!h]
  \includegraphics[width=11cm]{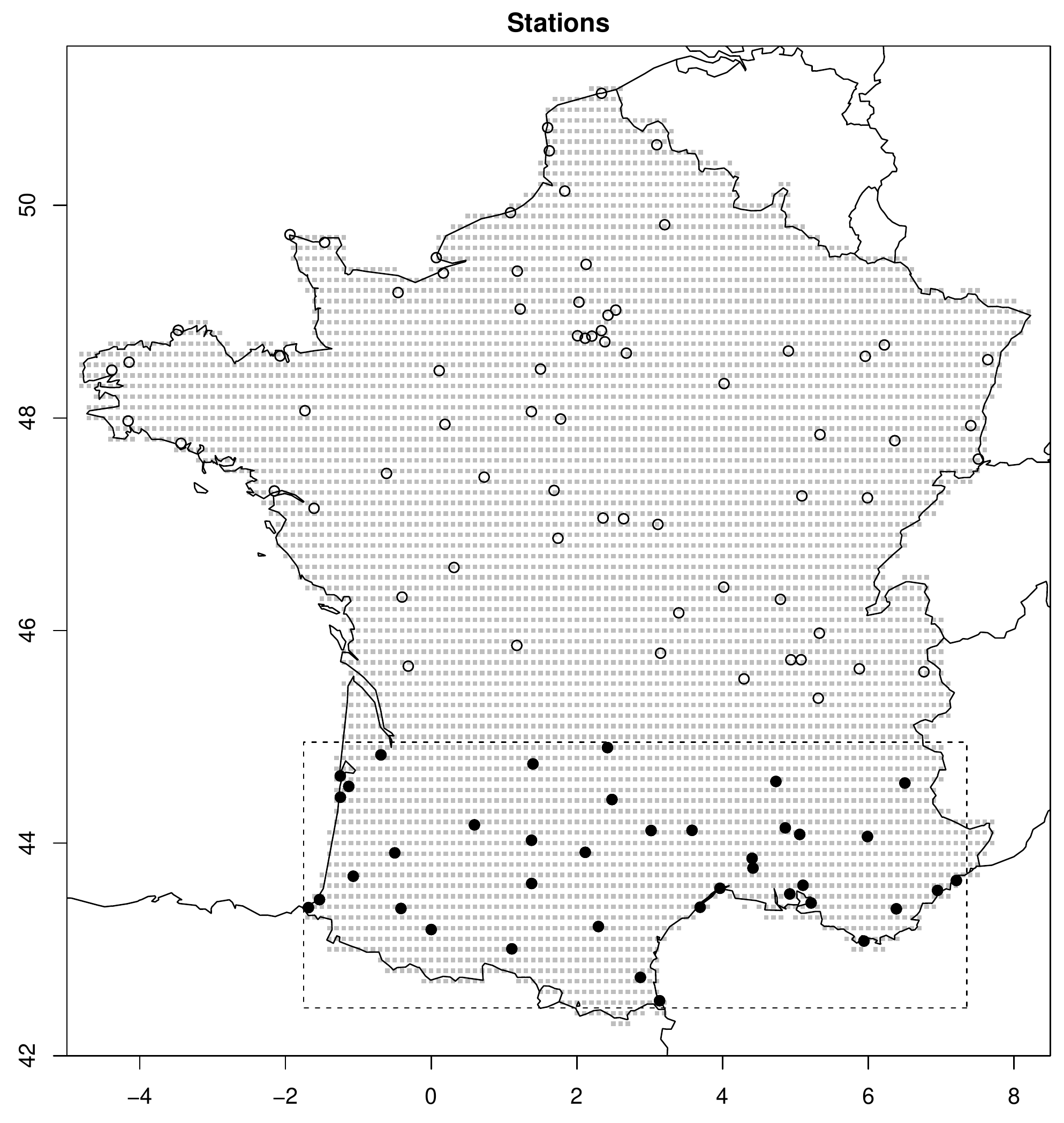} \includegraphics[width=11cm]{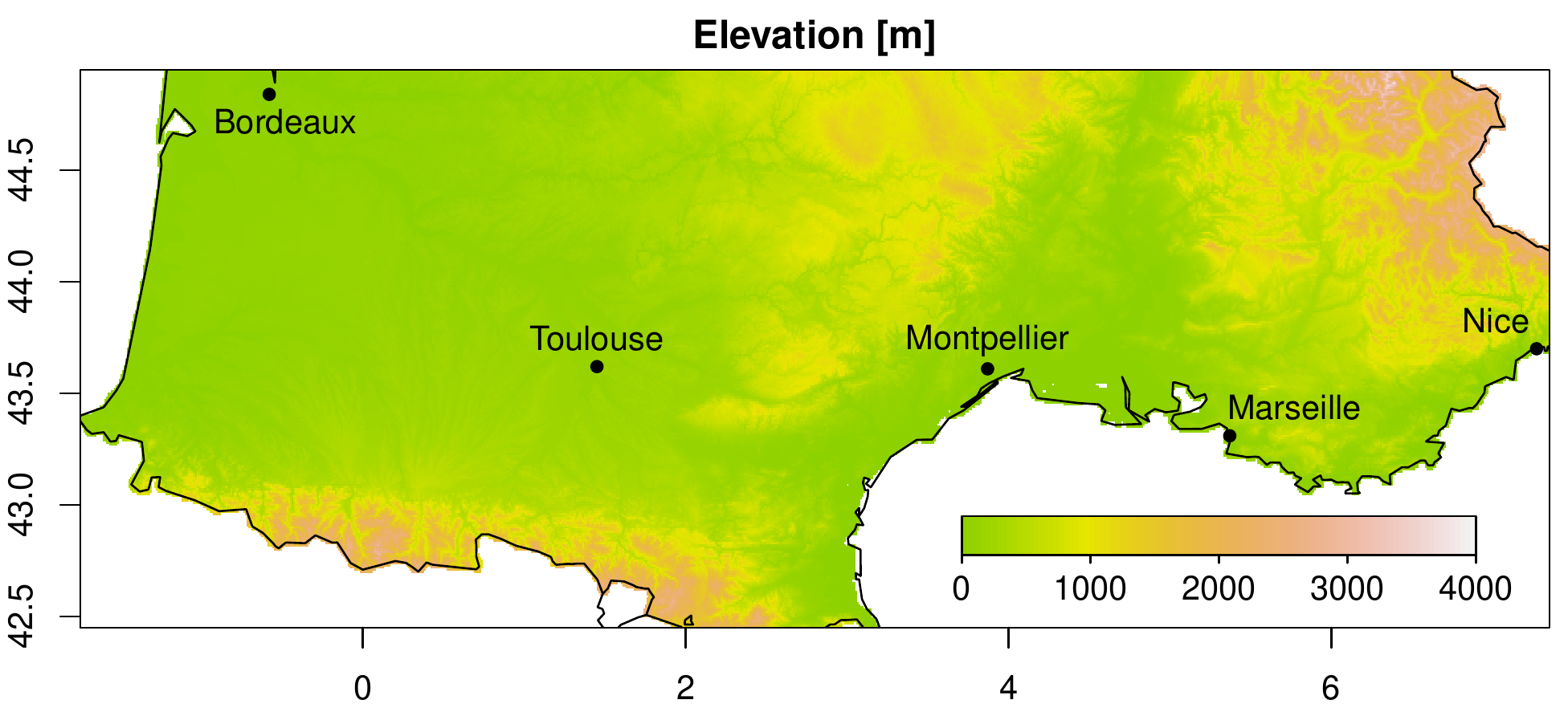}
  \caption{Top panel:  our 39 locations of interest   from the 110 reference M\'et\'eo France weather stations network; the finer and homogeneous grid in grey corresponds to PEARP  points of ensemble forecasts. 
  Bottom: altitude map of our region of interest to highlight the strong orography  from sea level to 4000 meters; the elevation data are retrieved from the Terrain Tiles on Amazon Web Services via the \texttt{R} package \texttt{elevatr} \citep{elevatr}.}
  \label{fig:grid}               
\end{figure}

The mainland French territory witnesses complex spatial rainfall weather patterns due to its changing  orography, the influence of the Atlantic Ocean, the Mediterranean Sea and different small and large climatological factors such as NOA. Another important aspect when modeling rainfall distributions is the quality of the data at hand. 
Here, our goal is to build our statistical analysis from the reference network of M\'et\'eo France that is composed of 110 well kept weather stations, see the white and black dots in Figure \ref{fig:grid} (top). These high quality stations recorded daily rainfall amounts over the time period  1980--2017.  
The elevation map in the bottom panel of Figure \ref{fig:grid} displays a strong orography over the south of France, see e.g.\ the C\`evennes region north of Montpellier, the Pyrenees in the southwest and the Alps in the east. These geographical features suggest that either anisotropy, non-stationarity or both can be expected to be present in the  spatial component of  heavy rainfall, even after removing spatial trends at the marginal levels. Autumnal moisture brought from the Mediterranean sea can lead to severe convective storms with a specific spatial structure, different from northern weather patterns stopped by the Pyrenees.

As seen in Section \ref{sec:intro-max}, max-stable statistical models can capture strong dependence among maxima, but they are not appropriate to model weak dependencies for extremes \citep[cf.][for instance]{wadsworth-tawn-12}. In contrast to temperature extremes like heat waves, heavy rainfall are not likely to be dependent over very large regions.
Precipitation extremes recorded at two stations more than 500 kilometers apart are likely to be independent. For this reason, we reduce our area of interest from the whole mainland territory to the southern part of France, see the black dots and the box in the top panel and the corresponding altitude map in the bottom panel of Figure \ref{fig:grid}. 
 We have chosen this particular region because most severe heavy rainfall events occur there. 
 
%%%%%%%%%%%%%%%%%%%%%%%%%%%%%%%%%%%%
%%%%%%%%%%%%%%%%%%%%%%%%%%%%%%%%%%%%

\subsection{Daily rainfall  recorded by weather stations}\label{sec: stations}
\begin{figure}[!h]
 \centering \includegraphics[width=11cm]{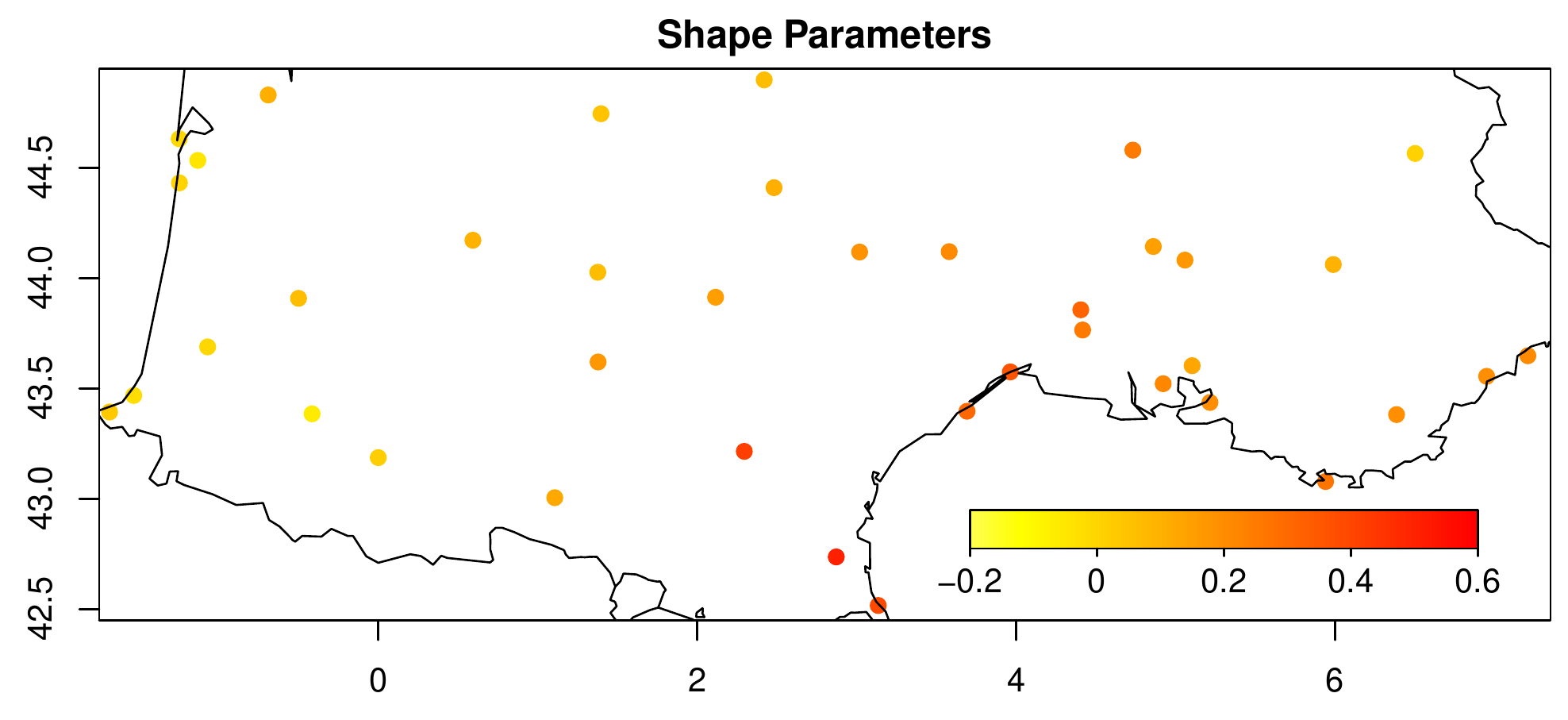}\\ \centering \includegraphics[width=11cm]{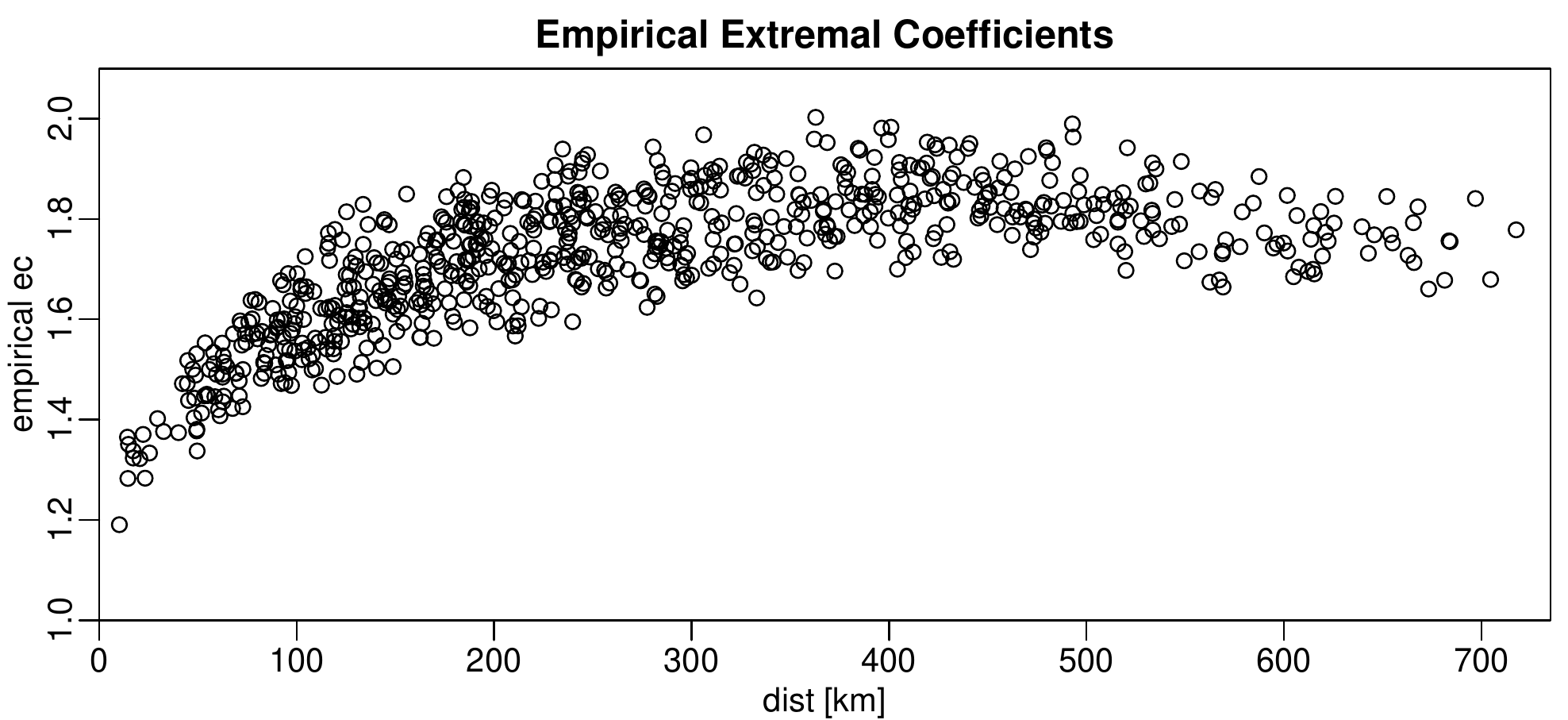}
 \caption{Top: Map of measurement stations and estimated GEV shape parameters; 
          Bottom: Pairwise empirical extremal coefficients, see Equation (\ref{eq:extr-coeff}),  plotted against the distance between the stations.}\label{fig:stations}
\end{figure}

We consider fall (SON) daily precipitation at $M=39$ stations in the south of France, see black dots in Figure \ref{fig:grid}, over 38 years, from 1980 to 2017. Each fall season has been divided into five blocks of length 18. Maxima were computed over each of these $5 \times 38 = 190$ blocks. The top panel of Figure \ref{fig:stations} displays the estimated GEV shape parameters obtained by a probability-weighted moment method \citep{DGNR08} separately for each station, assuming independence among blocks. Kolmogorov-Smirnov tests per stations indicate a good fit of the estimated GEV distributions with an average $p$-value of $0.591$. The value range and the spatial pattern of estimated $\xi$ is roughly similar to the ones observed in previous studies \cite[see, e.g.][]{CNN-17,CNAC-12,BMT-18}.

With the analysis of the marginal distributions ensuring the compatibility of the station data with a max-stable model, we will henceforth focus on our main objective, i.e.\  modeling the spatial structure among heavy rainfall. To this end, we first investigate the extremal dependence structure between rainfall data at different stations as described in Section \ref{sec:intro-max}. More precisely, for each pair of stations, we estimate the pairwise extremal coefficient via a rank-based empirical version of the weighted $F$-madogram, see \cite{marcon-etal-17}. In the bottom panel of Figure \ref{fig:stations}, the dependence captured by the estimated pairwise extremal coefficient decreases as the distance between stations increases. A nugget effect seems to be present because extremal coefficients for very small distances do not appear to be close to one, but rather around 1.2. 

The main task is to produce max-stable processes that can reproduce such spatial features at finer scales. To this end, we will make use of the second type of data, namely precipitation forecasts.

\subsection{Gridded daily precipitation produced from weather forecast center}\label{sec:ensemble} 
The national French weather service, M\'et\'eo France, produces, on a daily basis, an ensemble of 35 members with forecasted daily precipitation at 17596 cells of size $0.1^\circ \times 0.1^\circ$ covering the mainland of France, see the grid displayed in Figure \ref{fig:grid}. 
To merge these forecasted data  with our observational network described in Section \ref{sec: stations}, we extract a subset $S$ of $1752$ grid cells over a rectangular region that contains our $M=39$ stations (see the box in the bottom panel of Figure \ref{fig:grid}) and over a time period that overlaps, more precisely Fall seasons from 2012 to 2017, i.e.\ 546 days. Let $f_i^{(j)}(s)$ denote the forecast of the $j$th ensemble member for grid cell $s$ and day $i$.
In order avoid potential scaling problems, for each ensemble member $j \in \{1,\ldots,25\}$ and each grid cell $s \in S$, all the forecasts are transformed to a unit Fr\'echet scale via a rank transform 
$$ w_i^{(j)}(s) = \begin{cases} -1/\log(\mathrm{rank}(f_i(s))/(N+1)), & \quad f^{(j)}_i(s) > 0,\\
0, & \quad f^{(j)}_i(s) = 0, \end{cases} $$
where $N^{(j)}(s)$ denotes the number of days for which the forecast of ensemble member
$j$ is available at grid cell $s$. 
This results in $N = 17626$ transformed forecast maps. As we will treat all the ensemble members in the same way, henceforth, for simplicity, we will denote these $N$ maps  $w_1, \ldots, w_N$, with $w_i=(w_i(s))_{s \in S}$, of transformed forecasted daily precipitation.
These forecasts are used to construct data driven max-stable models for extreme precipitation.

%%%%%%%%%%%%%%%%%%%%%%%%%%%%%%%%%%%%
\section{Statistical models}\label{sec:models} 
\begin{figure}[!h]
	\centering
	\includegraphics[width=12cm]{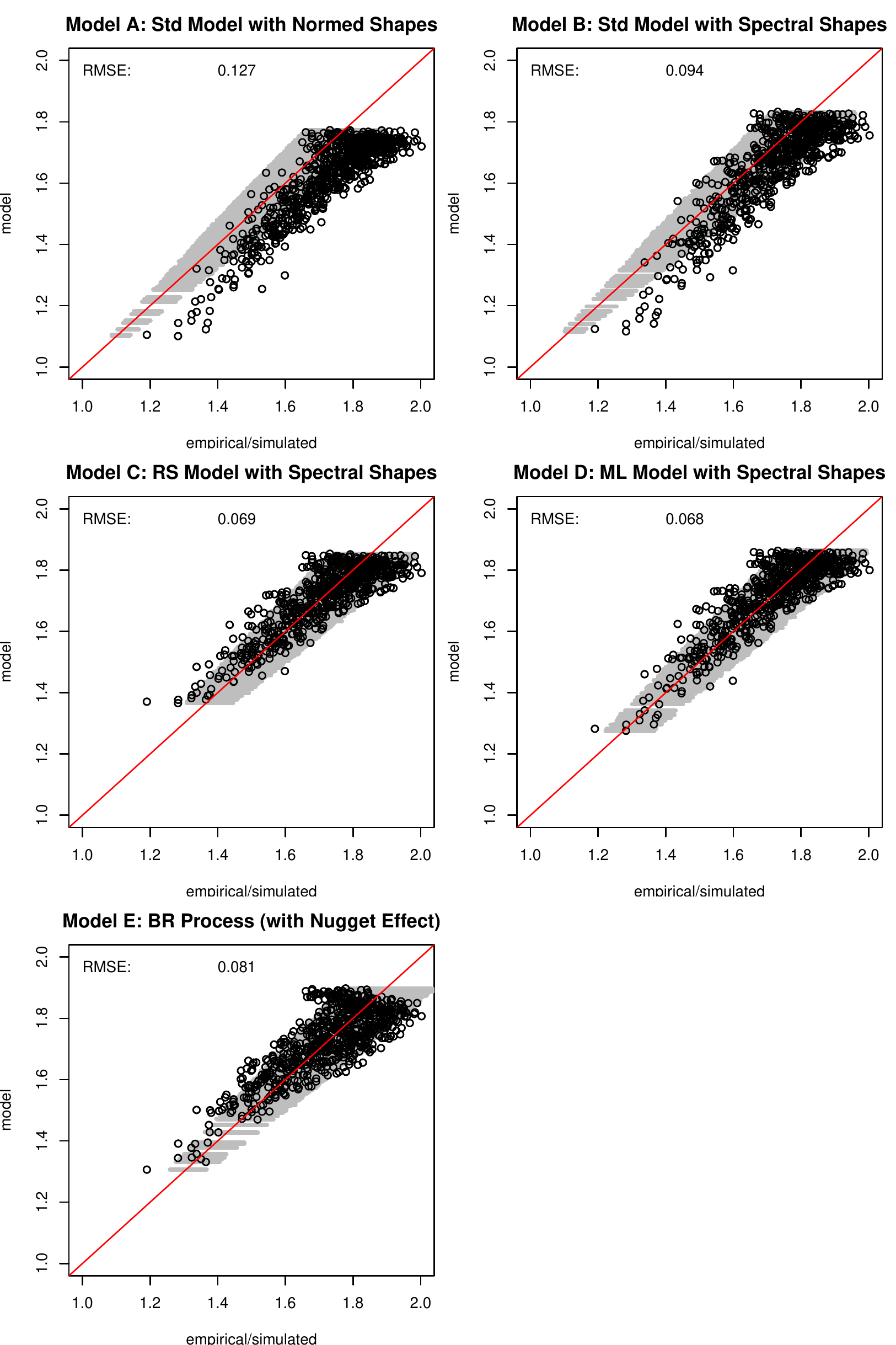}
	\caption{For each of the fitted models, Model A--E, the theoretical pairwise extremal
		coefficients are plotted against the empirical estimates (black); the gray intervals indicate the uncertainty of the estimates based on simulations from the fitted models.}
	\label{fig:all_ec_simu}
\end{figure}

%%%%%%%%%%%%%%%%%%%%%%%%%%%%%%%%%%%%
In this section, five different max-stable models are studied and compared. 
For simplicity and analogously to Section \ref{sec:intro-max}, all models are defined in a standardized way with unit Fr\'echet margins. While one of the five max-stable models is fully parametric and based on a Brown--Resnick process with a nugget effect, the other four will be data driven by the $N$ maps from Section \ref{sec:ensemble}. 
To assess all five models, we will focus on their extremal dependence structure. 
In particular, the extremal coefficients associated to each model will be compared to  their empirically estimated counterparts in Figure \ref{fig:all_ec_simu}. In this comparison, each station is spatially identified to its closest grid cell. The resulting root mean squared error (RMSE) will be used to evaluate the quality of the model fit.
Estimation uncertainty will be assessed by a parametric bootstrap. 
More precisely, 190 block maxima are simulated from each of the fitted models 500 times. 
For each of these 500 simulations, the pairwise extremal coefficients are estimated. The intervals between empirical $2.5\,\%$- and $97.5\,\%$-quantiles of these samples are displayed in gray, indicating the region in which the empirical estimates would be expected to be if the fitted model was correct. 

The five models are listed below. 
\begin{itemize}
  \item {\bf Model A}: As a starting point, we assume that all $N$ maps contain relevant information about rainfall maxima. 
  Hence, the standard max-linear model \eqref{eq:maxlin} is implemented by  defining   
  $$ 
  z_i(s) = \frac{w_i(s)}{\frac 1 N \sum_{j=1}^N w_j(s)}, \qquad i=1,\ldots,N,
      \ s \in S,
      $$   
      as normalized basis functions based on all the forecast vectors $w_i$. 
\end{itemize}
According to the top left panel of Figure \ref{fig:all_ec_simu}, Model A does not capture accurately the extremal dependence structure observed in biweekly maxima recorded at weather stations. 
This shortcoming can be explained by the incorrect assumption that all grid points of all  daily fields are linked to extreme rainfall. Days with little or no rain, however, should be not used to build the basis functions. To account for this fact, we exploit the theory of generalized Pareto processes \citep{dehaan-ferreira-14}. According to the theory, extremal dependence in the forecasts should fully described by the max-spectral functions $w_i / \|w_i\|_\infty$ for those $i$ such that $\|w_i\| \geq u$ for some high threshold $u$. Thus, we will use these spectral functions to build new basis functions. Assuming, without loss of generality, that the vectors are sorted w.r.t.\ their maximum, i.e.\ $\|w_1\|_\infty \geq |w_2\|_\infty \geq \ldots \geq \|w_N\|_\infty$, this results in a number $N(u)$ of forecasts $w_1, \ldots, w_{N(u)}$ to be taken into account. 
For simplicity, for each of the following models, Model B, C and D, we will choose a fixed threshold $u$ as the empirical $90\,\%$-quantile of the vector $(\|w_i\|_\infty)_{i=1,\ldots,N}$. This choice leads to $N(u) = 1819$ maps used for the construction of the models.

\begin{itemize}
	\item {\bf Model B}: As an improvement  of Model A, we consider a max-linear model with max-spectral functions built from those forecast exceeding the threshold $u$. The resulting normalized basis functions are given by
	$$ z_i(s) = \frac{ w_i(s) / \|w_i\|_\infty}{\frac 1 N \sum_{j=1}^{N(u)} w_j(s) / \|w_j\|_\infty}, \qquad i=1,\ldots, N(u).$$ 
\end{itemize}

The top right panel in Figure \ref{fig:all_ec_simu} indicates that the extremal coefficients for pairs of strongly dependent stations are still systematically smaller than their empirical counterparts, i.e.\ dependence in the model is stronger than dependence in the data -- a phenomenon that is often present when comparing forecasts to observations and 
that can be explained by the presence of some nugget effect in the observed data. Models C and D provide two ways of incorporating nugget effects into
max-stable models. 
\begin{itemize}
  \item {\bf Model C}: This model is based on \cite{reich-shaby-12} construction defined
        by \eqref{eq:reich-shaby}. The  spectral functions are identical to the ones used in Model B.
  \item {\bf Model D}: To handle a possible nugget effect, we combine Model B with a noise
        process throughout a max-linear operator 
        $$ Y(s) = \max\{aY_{\mathrm{noise}}(s), (1-a) Y_{\mathrm{B}}(s)\} , \qquad s \in S, $$
        where $a \in [0,1]$ is a mixture parameter, $Y_{\mathrm{noise}}$ is a noise process with unit Fr\'echet marginal distributions and, independently of $Y_{\mathrm{noise}}$, $Y_{\mathrm{B}}$ is the max-linear process in Model B.
\end{itemize} 
The additional parameters, $\alpha \in (0,1)$ in Model C and 
$a \in [0,1]$ in Model D, respectively, are chosen such that the RMSE is minimized. 
By a first visual inspection, see Figure \ref{fig:all_ec_simu}, models C and D appear to capture the observed extremal coefficients well. 

The quality of these plots has to be interpreted in regard to the number of parameters inferred. This number is zero, one and one for model B, C and D, respectively. This highlights that our approach is very parsimonious in terms of parameter number and inference complexity. 
The threshold  was  set to   the  $90\,\%$-quantile of $(\|w_i\|_\infty)_{i=1,\ldots,N}$ for all models, and consequently, the RMSE could be even lower if the 
threshold choice was optimized for each of the three models separately. 
But, as our goal is to propose a straightforward estimation scheme, we refrain from  optimizing the choice of individual thresholds for models B, C and D.

As previously mentioned, our last model is different and based on a classical and fully
parametric max-stable model, the Brown--Resnick process. 
 \begin{itemize}
	\item {\bf Model E}: We use a Brown--Resnick process associated to the variogram
	$$ \gamma(h) = \sigma^2 \mathbf{1}_{\{\|h\|= 0\}} + \left\| \begin{pmatrix} b_1 & 0 \\ 0 & b_2 \end{pmatrix} 
	\begin{pmatrix} \cos(\theta) & \sin(\theta) \\ -\sin(\theta) & \cos(\theta) \end{pmatrix} h \right\|^\beta, \quad h \in \RR^2,$$
	for some $\sigma^2>0$, $b_1, b_2 > 0$, $\theta \in (-\pi/4, \pi/4)$, $\beta \in (0,2]$. Here, the five parameters are chosen such that the RMSE of the estimated pairwise extremal coefficients is minimized.
\end{itemize}
By construction, this model cannot capture non-stationarity and increasing the number of parameters to do so will be non-trivial. 
Still, Model E offers some flexibility in terms of anisotropy and nugget effects.

\medskip

In the following, we will compare the classical Model E to the two data driven models C and D which provide the best fit in terms of pairwise extremal coefficients as indicated by the root mean squared errors displayed in Figure \ref{fig:all_ec_simu}. It can be seen that the theoretical extremal coefficients according to models C and D are almost identical apart from the coefficients for the strongly dependent pairs where dependence seems to be slightly stronger in Model D than in Model C.
The coefficients according to Model E are also largely similar to the ones obtained by the data driven models. The main difference is that the fitted model E exhibits slightly larger extremal coefficients than the empirical ones for strongly dependent pairs, while the theoretical coefficients are slightly smaller than their empirical counterparts for weakly dependent pairs of stations. These inaccuracies are also reflected by the root mean squared error for model E. The relative gain from model E to model D is 16\% in terms of RMSE.
 Analogously to the pairwise dependence structures, one may also compare the fits for summary statistics of higher order. The corresponding results for the triplewise extremal coefficients can be found in the appendix.

\begin{figure}[!h]
 \centering 
 \includegraphics[width=14.5cm]{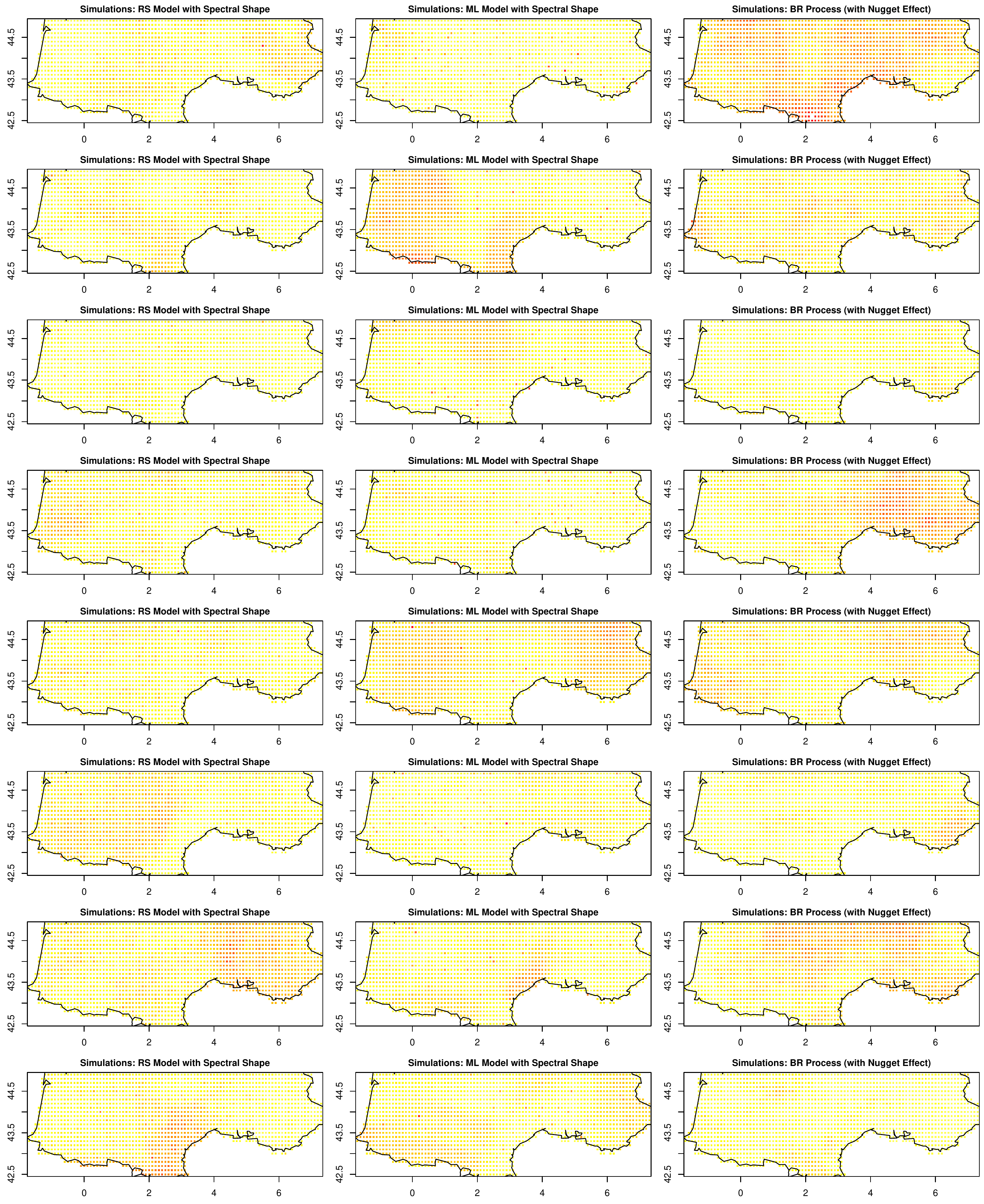} 
 \caption{Simulations from Models C, D and E.}\label{fig:simu}
\end{figure}

While the summary statistics considered so far provide some information about extremal dependence along the diagonal of bivariate and trivariate distributions, respectively, some further insight in the three models may be gained by regarding artificial precipitation fields obtained by simulations. Such realizations from the three models are displayed in Figure \ref{fig:simu}. Although such graphical comparisons are only qualitative, models C and D appear to be able provide a wide range of spatial features with specific regional behavior,
having a slightly stronger nugget effect than Model E.
They can reproduce very localized events, but also generate simulated fields that appear to have well defined spatial structures along geographical features, see the last two rows. 

The stationary model E with five parameters appear to simulate spatial structures similar to the simpler models C and D with one parameter only. This fact leads to less expensive inference schemes for the data driven models. Furthermore, due to their simple structure as given in Equations \eqref{eq:maxlin} and \eqref{eq:reich-shaby}, respectively, the data driven models are easy to simulate, while simulation of Brown--Resnick or other parametric spatial max-stable models on dense grids is typically computationally intensive
\citep[cf.][for instance]{DEO16,OSZ18}.

%%%%%%%%%%%%%%%%%%%%%%%%%%%%%%%%%%%%
\section{Discussion and conclusion} \label{sec:conclusion}
%%%%%%%%%%%%%%%%%%%%%%%%%%%%%%%%%%%%
Although climatologists, weather forecasters and statisticians have been collaborating extensively the last decades, the field of data assimilation being a successful  example of such a joint research effort, the extreme value theory community has been slower at integrating new data sources within their multivariate  extremal models. 
For example, there are many high quality methodological articles on heavy rainfall analysis, but most are based on  complex parametric models applied to one unique data source. 
This leads to non-trivial inference problems and such approaches can be difficult to transfer to researchers  outside of this particular domain.

In this work, our goal was to show that the framework of max-stable processes, one pillar of spatial extreme value analysis, can be easily coupled with other data sources. More specifically, simple max-stable processes that integrate ensemble forecast rainfall data as spectral profiles were able to reproduce the main spatial features of heavy rainfall over a complex climatological region. We also show that our approach compares favorably with a  more complicated parametrized model. In addition, it is simple to  handle non-stationarity,
and both inference and simulation are straightforward and quick. One obvious limitation of our method is that, as expected from the theory of max-stable processes, the efficiency of our approach directly depends  on the quality of the input data. If weather services were unable to reproduce adequately  important spatial features of storms and fronts in their forecast ensembles, our strategy will naturally be inefficient. 

To conclude, the production of numerical models outputs, their capabilities, their associated resolution and their sizes appear to have increased rapidly these last few years, and this trend is likely to continue. 
In the context of extreme value analysis, it is always a delicate question to know if such numerical models can simulate adequately extreme events, or produce even unobserved ones. 
From a geophysical point of view, most of these numerical models are physically consistent, and consequently should contain some meaningful information about spatial and temporal structures. 
Besides the case study presented in this paper, it would be interesting to determine if other extremal models, e.g. asymptotic independence, could benefit of such additional information to improve the estimation of very high quantiles, and if so, how to combine them to produce spatio-temporal extremal fields in compliance with EVT in a physically coherent way.

%%%%%%%%%%%%%%%%%%%%%%%%%%%%%%%%%%%%
\section*{Acknowledgments}
We would like to thank M\'et\'eo France for providing the data used in this work. 
Part of P.~Naveau's work was supported by the European DAMOCLES-COST-ACTION on compound events, and also benefited from  French national programs, in particular FRAISE-LEFE/INSU,  MELODY-ANR, and ANR-11-IDEX-0004 -17-EURE-0006.

\bibliographystyle{imsart-nameyear}

\begin{appendix}
\section*{Trivariate dependence structure}

While pairwise extremal coefficients are used for model fitting, we can also consider 
extremal coefficients of higher order. Analogously to Equation \eqref{eq:def-ec}, for 
a max-stable process $Y$ with marginal distribution $G$ and $s_1,\ldots,s_d \in S$, the extremal coefficient $\theta(s_1,\ldots,s_d)$ can be defined via
\begin{equation*}
\PP( G(Y(s_1)) \leq u, \ldots, G(Y(s_d)) \leq u) 
= \PP( G(Y(s_1)) \leq u)^{\theta(s_1,\ldots,s_d)}, \qquad u \in [0,1].
\end{equation*} 
By construction, $\theta(s_1,\ldots,s_d) \in [1,d]$ -- a quantity that is often interpreted as the number of independent random variables among $Y(s_1),\ldots,Y(s_d)$.
Likewise the pairwise coefficients, the higher order coefficients can be estimated via the empirical multivariate weighted $F$-madogram, see \citealp{marcon-etal-17}).

\begin{figure}[!h]
	\centering \includegraphics[width=6cm]{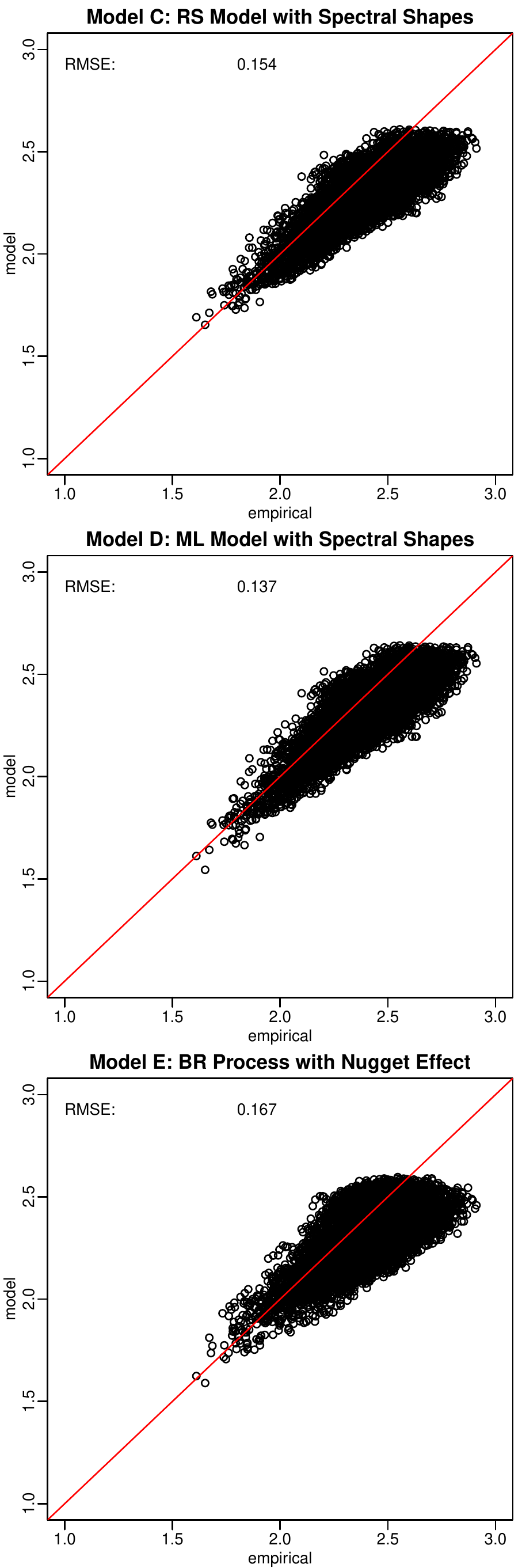} 
	\centering \includegraphics[width=6cm]{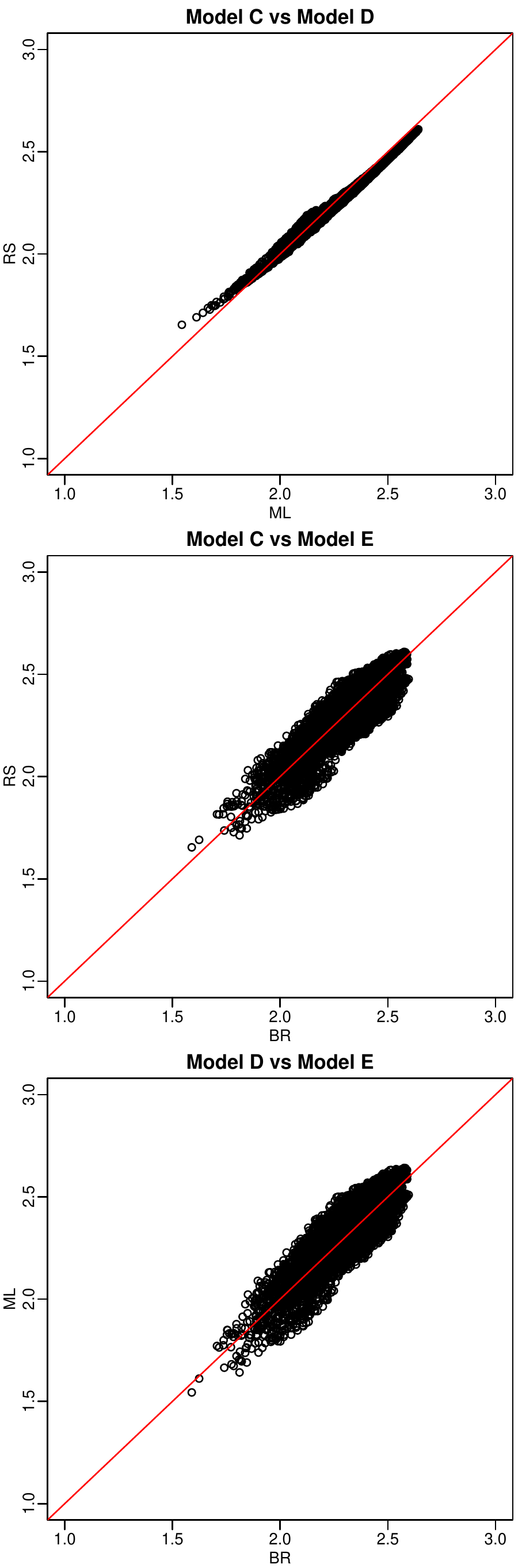}
	\caption{Triplewise extremal coefficients for three fitted models, Model C, 
		Model D and Model E.
		Left: analogously to Figure \ref{fig:all_ec_simu}. 
		Right: Comparison of the
		three models by plotting the theoretical triplewise extremal
		coefficients against one another.}
		\label{fig:triple_ec}
\end{figure}

In Figure \ref{fig:triple_ec}, results for the estimated triplewise extremal coefficients are shown and compared to the theoretical ones for the fitted models. Similarly to the results for the pairwise coefficients displayed in Figure \ref{fig:all_ec_simu}, the root mean square error for Model E is slightly worse than the errors for Model C and Model D, respectively. Furthermore, it can be seen that the triplewise extremal coefficients of Models C and D are nearly identical, while the coefficients of Model E show stronger deviations.

\end{appendix}

\end{document}